\documentclass[twocolumn,floats,rmp]{revtex4}

\usepackage{amsmath}
\usepackage{graphicx}

\begin{document}

\title{Lateral--Pressure Profiles in Cholesterol--DPPC Bilayers}

\author{Michael Patra}
\affiliation{Biophysics and Statistical Mechanics Group,
Laboratory of Computational Engineering, \\
Helsinki University
of Technology, P.\,O. Box 9203, FIN--02015 HUT, Finland}
\affiliation{Physical Chemistry 1, Centre for Chemistry and
Chemical Engineering, \\
Lund University, P.\,O. Box 124, SE--22100 Lund, Sweden}

\begin{abstract} 

By means of atomistic molecular dynamics simulations, we study 
cholesterol--DPPC (dipalmitoyl phosphatidylcholine) bilayers of different
composition, from pure DPPC bilayers to a 1:1 mixture of DPPC and cholesterol.
The lateral-pressure profiles through the bilayers are computed and separated
into contributions from the different components. We find that the pressure
inside the bilayer changes qualitatively for cholesterol concentrations of about
$20\,\%$ or higher. The pressure profile then turns
from a rather flat shape into an alternating sequence of  regions with large
positive and negative lateral pressure. The changes in the lateral-pressure
profile are so characteristic that specific interaction between
cholesterol and molecules such as
membrane proteins mediated solely via the lateral-pressure
profile might become possible.

\end{abstract}

\maketitle

\section{Introduction}

From a macroscopic point of view a planar lipid membrane in equilibrium has,
by definition, a vanishing surface tension. However, on a microscopic
level there is, within the membrane, local lateral pressure, i.\,e., pressure
tangential to the interface. Only when all of the local contributions are summed
and averaged along the bilayer normal, the net pressure vanishes. The local
lateral pressure arises from the different structural components of the lipid
bilayer~\cite{marsh:96a} and, almost counter-intuitively, each of these
contributions can be of the order of several hundreds of bars.

The importance of lateral pressure profiles has been discussed in several recent
reviews~\cite{ben:95a,marsh:96a,kinnunen:00a,bezrukov:00a,eckenhoff:01a}. The
increased interest in understanding lateral pressure profiles is easily
understood as lateral pressure has been proposed to have an important role in,
for example, general
anaesthesia~\cite{cantor:97b,eckenhoff:01a} or
inhibition and regulation of protein function~\cite{kruijff:97a}
(see references 25--50 in Ref.~\onlinecite{brink:04a} for a list
of proteins for which a relation to lateral pressure has been suggested).

Due to the absence of good probes for lateral pressure, direct experimental
measurements are difficult and only a single experimental study exists at the
present~\cite{templer:98a}. In contrast, computer simulations have for the past
ten years been able to supply direct, yet not straightforward, access to study
pressure profiles and their response to changes in the membrane. Calculations of
lateral pressure profiles are a delicate matter as already the introduction of
small simplifications to the system can render the results questionable. For
example, the reported results for pure lipid bilayer systems from coarse-grained
simulations~\cite{goetz:98a,harries:97a,shillcock:02a} disagree with their
counterparts from atomistic simulations~\cite{lindahl:00b,gullingsrud:04a}.

In this paper, we study the effects of cholesterol on lateral pressure.
Cholesterol is an essential component of all Eukaryotic cell membranes
where it plays a crucial role for both static structure and
dynamics~\cite{yeagle:85a,simons:00a}. In particular it regulates the fluidity
of the cell membrane~\cite{mcmullen:96a}. All of this goes along with
changes in the lateral pressure profile. Indirect evidence for
the importance of cholesterol on the lateral pressure comes from studies of
membrane channels which are highly sensitive to the pressure of their
environment~\cite{suhkarev:97a,hamill:01a}. One such channel is the
nicotinic acetylcholine receptor which ceases to function in the absence of
cholesterol~\cite{rankin:97a}.

Apart from the fundamental importance of cholesterol and its effect on lateral
pressure, the study of cholesterol is interesting for a second reason:
cholesterol is a highly specific molecule. Already small modifications of its
sterol structure lead to significant changes of the membrane
properties~\cite{endress:02a,scheidt:03a}. Generic theories of lateral pressure
describe a molecule basically only by its volume and its rigidity. When such
theories are applied to cholesterol~\cite{cantor:99a}, they thus inevitably 
fail to capture many essential features of cholesterol, and even
predict pressure changes of wrong sign in some parts of the bilayer.

In this article, we thus use atomistic molecular dynamics simulations to study
lateral pressure profiles for six different systems, ranging from a pure DPPC
bilayer to a bilayer consisting of a 1:1 mixture of DPPC and cholesterol. This
paper is, to the author's knowledge, the first detailed atomistic computational
study addressing the effect of cholesterol or other small molecules on the
build-up of the lateral pressure profile in phospholipid membranes.

\section{Local pressure}

The pressure tensor $\tensor{p}$ can be computed as
\begin{equation}
	\tensor{p}=2\tensor{E}-\tensor{\Sigma}\;,
	\label{eqDruck}
\end{equation}
from the kinetic energy density tensor $\tensor{E}$ and the configuration
stress tensor $\tensor{\Sigma}$. Both of the latter quantities can be expressed
in terms of atomistic positions, velocities and forces as
\begin{gather}
	\tensor{E}=\frac{1}{2}\sum_i m_i \vec{v}_i \otimes \vec{v}_i \;, 
		\label{eqEnergie}\\
	\tensor{\Sigma}=\frac{1}{V} \sum_{i<j} \vec{F}_{ij} \otimes
		\vec{r}_{ij} \;,
		\label{eqKraft}
\end{gather}
and are thus accessible in a MD simulation.
While the above expressions are, strictly speaking, defined only if the 
summations are extended over the entire simulation volume~\cite{heinz:03a}, in 
practise it is possible to divide the different contributions into slices 
according to the positions of the involved atoms~\cite{lindahl:00b}.

This implies, however, that the force $\vec{F}_{ij}$ between particles $i$ and
$j$ is known explicitly. This is not the case if a multipole or lattice based 
method (such as PME) is used to evaluate electrostatic interaction. In the first
reported atomistic computation of pressure profiles~\cite{lindahl:00b}
electrostatics was therefore truncated a distance of $1.8~\mathrm{nm}$. As it
is known by now, however, using abrupt truncation, especially at such a short
distance,
introduces significant artifacts into bilayer
systems~\cite{patra:03b,anezo:03a,patra:04b,patra:04c} and thus needs to be
avoided. We thus use reaction-field technique that has been
shown to give results consistent with the application of long-range
electrostatics~\cite{patra:04c} while at the same time employing explicit
expressions for $\vec{F}_{ij}$.

The global pressure is equal to the average of the local pressures, and any
condition on the global pressure thus translates onto the pressure profile. 
Since the outside of a bilayer is at equilibrium with the environment (i.\,e.,
approximately $1~\mathrm{bar}$ in most cases), the average local pressure
has to be equal to that value. If the global pressure would be different,
the system would react by shrinking or expanding, and thus would not be in
equilibrium. 

Still, there is nonvanishing local pressure even in equilibrium. The existence
of an interface between the water and the lipid goes along with an energy
penalty which could be lowered by packing the bilayer more densely, thereby
decreasing the area per lipid. Steric constraints between the lipid tails
prevent this from happening. The equilibrium value of the area per lipid is thus
a compromise between the ``wishes'' of the head groups and the tails. This is
directly reflected in the lateral--pressure profile. Since the interface region
prefers a further reduction of the area per lipid, the local lateral pressure
there is negative (pointing inwards) whereas it is positive in the tail region
(pointing outwards).

The lateral pressure profile in equilibrium thus is a direct reflection of the
inhomogeneity of the bilayer along the bilayer normal. In
contrast, the bilayer is homogeneous parallel to the bilayer interface. In
equilibrium, the normal component of the local pressure thus has to vanish
everywhere. (More correctly, it has to be constant and equal to the applied
external pressure of $1~\mathrm{bar}$.)
In contrast to the above theoretical argument, a non-vanishing normal
pressure component is found in the numerical simulations, in magnitude about
$10\,\%$ of the lateral component. We attribute this to the use of
distance-constraints in our simulation. For computational efficiency, the
distance between bonded atoms is kept constant, meaning that the system cannot
locally expand or contract to reduce the local pressure. This ``constraint
force'' was studied in Ref.~\onlinecite{lindahl:00b}, and the values quoted
there are able to explain the values of the normal pressure found in our
simulation. Since effects of numerics are isotropic on average, we can thus
improve the lateral component of the pressure by subtracting the normal
component from it before the analysis. This will be done throughout this paper.

\section{Simulation details}

We study lipid bilayers comprised of 128 molecules (64 per leaflet), at various
ratios of dipalmitoyl
phosphatidylcholine (DPPC) 
and cholesterol, hydrated by
3655 water molecules. DPPC molecules are described by the model from
Ref.~\onlinecite{tieleman:96a}, which utilises
the description of lipids from Ref.~\onlinecite{berger:97a}. 
Cholesterol was described by the model from Ref.~\onlinecite{hoeltje:01a}
and the SPC model~\cite{berendsen:81a} was used to describe water.
The simulations were performed using the Gromacs package, both in the standard
release~\cite{lindahl:01a} and in an adapted version that allows the computation
of local pressures~\cite{lindahl:00b}.

For computing electrostatics interactions, we employed a twin-range
setup~\cite{bishop:97a} in which the interactions within a distance
$r_{\mathrm{list}}=1.0~\mathrm{nm}$ were evaluated at every integration step, 
and those between $r_{\mathrm{list}}$ and $r_{\mathrm{cut}}=2.0~\mathrm{nm}$
only every tenth integration time step. A reaction-field 
approach~\cite{tironi:95a} was used to account for interaction outside of
$r_{\mathrm{cut}}$ by assuming a homogeneous dielectric with $\epsilon=80$.
Lennard--Jones interaction was truncated at $1.0~\mathrm{nm}$.

DPPC, cholesterol and water molecules were
separately coupled to a heat bath at temperature $T = 323~\mathrm{K}$, and the
pressure was kept at $1~\mathrm{bar}$, both using the Berendsen
algorithms~\cite{berendsen:84a}. The size of the simulation box in the plane of
the bilayer ($x$-$y$ plane) was allowed to fluctuate independently of its
height.

As initial configurations for all simulations we used the final configurations 
of $100~\mathrm{ns}$ simulations~\cite{falck:04a} where electrostatics were 
treated by particle-mesh Ewald (PME)~\cite{essman:95a,frenkel:02}. 
The actual simulation procedure was divided into two steps. 
In a first step, $50~\mathrm{ns}$ trajectories were generated from these
structures for all
cholesterol concentrations using the standard version of Gromacs.
Electrostatics was treated by reaction field technique~\cite{tironi:95a},
and the generated
configurations were saved every $10~\mathrm{ps}$.
The bond lengths of DPPC molecules were constrained by the LINCS
algorithm~\cite{hess:97a} and water molecules were kept rigid by the SETTLE
algorithm~\cite{miyamoto:92a}, such that an integrator time step of
$2.0~\mathrm{fs}$ could be used. In the second step, the pressure profiles were
generated from this trajectory using an adapted version of
Gromacs (see below for details). In these runs, the 
SHAKE algorithm~\cite{ryckaert:77a} was used to constrain bond lengths

\begin{figure}[b]
\centering
\includegraphics[width=\columnwidth]{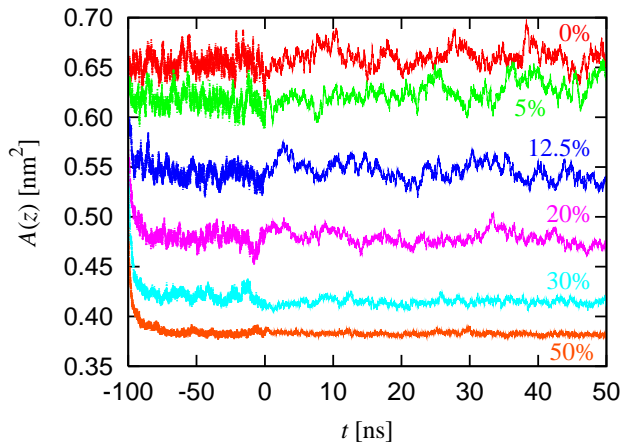}
\caption{Temporal development of the area per lipid for the different cholesterol
concentrations from $t=0~\mathrm{ns}$ to $t=50~\mathrm{ns}$. The results computed 
using PME are shown in condensed form from $t=-100~\mathrm{ns}$ to
$t=0~\mathrm{ns}$.}
\label{figArea}
\end{figure}

Resolving the pressure spatially makes the simulations about one order of
magnitude slower than a normal simulation. The pressure profiles were thus
computed in a second step, based on the saved configurations. Starting at every
saved configuration, a $4~\mathrm{ps}$ simulation was run to compute the
pressure profile. The first $2~\mathrm{ps}$ were ignored to rule out effects of
finite precision of the saved configurations. Unless mentioned otherwise, only
the final $30~\mathrm{ns}$ of each simulation were included in the analysis.

The instantaneous pressure $p$, computed from Eq.~(\ref{eqDruck}), is
fluctuating quickly in time. Even when the instantaneous spatial average over
the entire simulation box is considered, the pressure easily changes by several
hundred bars within a single integration time step. Computing a statistically
relevant pressure profiles thus is numerically challenging since one has to
sample a large number of configurations. We evaluated the pressure profile for a
number of simulation frames that is far larger than in previous studies where
pressure profiles of bilayers were computed~\cite{lindahl:00b,gullingsrud:04a}. 
This allowed us to divide the simulation box into $150$ bins for computing the
pressure profile, with the result having only negligible numerical noise. The
remaining uncertainty in the pressure profile is mainly due to the temporal
change of the area per lipid. To arrive at this data quality, a total of
approximately 25\,000 hours of cpu time was needed.


In previous MD studies on cholesterol--DPPC bilayers, electrostatics were 
handled either by plain cutoff~\cite{tu:98a} or
PME~\cite{hofsaess:03a,falck:04a}. In our simulations we used reaction-field
technique, motivated by the mutual consistency between reaction-field technique
and PME found for pure DPPC systems~\cite{patra:04c}. From our simulations we
found that the results with reaction-field technique and PME are almost
identical also for mixed cholesterol--DPPC bilayers. We thus refrain from
reproducing the entire standard set of quantities that are used to characterise
a bilayer. Rather, we show only the temporal development of the area per lipid
in Fig.~\ref{figArea}. The results of the earlier simulations~\cite{falck:04a}
done using PME are also shown there in compressed form from $t=-100~\mathrm{ns}$
to $t=0~\mathrm{ns}$. From Fig.~\ref{figArea} it is then immediately obvious
that changing the electrostatics treatment from PME to reaction field at
$t=0~\mathrm{ns}$ has no relevant effect onto the systems, and that the systems
are in equilibrium.

\section{Pressure profiles}

\begin{figure}
\centering
\includegraphics[width=\columnwidth]{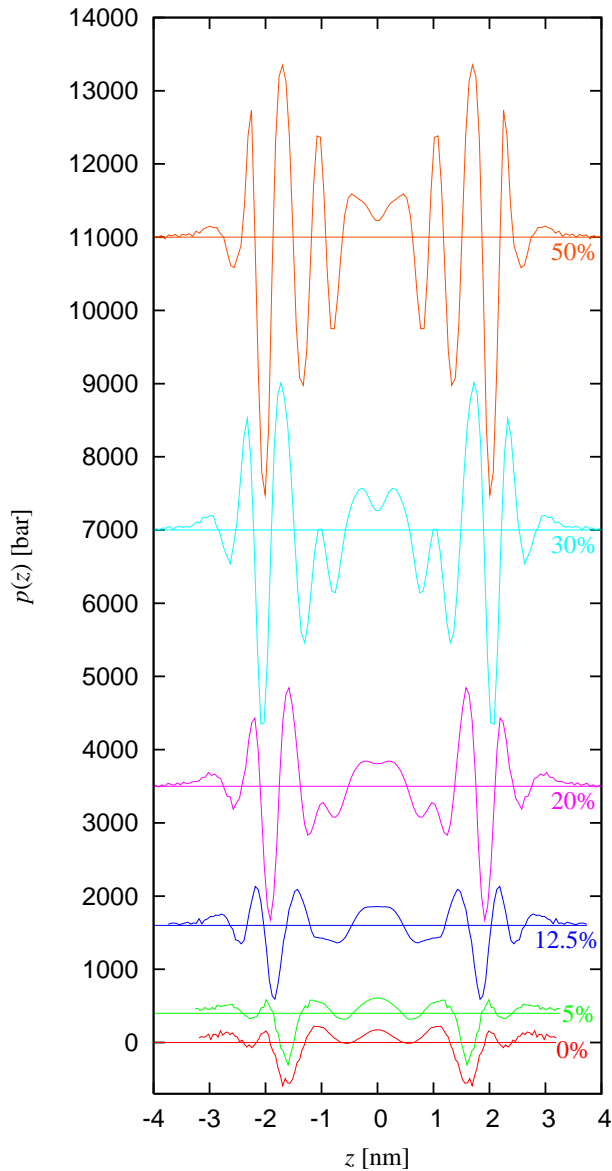}
\caption{Lateral pressure profiles through the bilayer for different cholesterol 
concentration. The different curves have been shifted for clarity. The labels on
the graphs give the cholesterol concentration.}
\label{figGlobDruck}
\end{figure}

The computed lateral pressure profiles are summarised in
Fig.~\ref{figGlobDruck}. We first want to discuss a few general features.

First, for low cholesterol concentration, the pressure does not decrease to zero
at the edges of the simulation box. This is a sign that the bilayer is not fully
hydrated, and it was estimated that an additional $4$--$5$ water molecules per
lipid would be needed for complete hydration~\cite{lindahl:00b}. This number is
in agreement with our results. Increasing cholesterol concentration means a
decrease of the number of lipid molecules, hence an increase in the number of
water molecules per lipid. At $12.5\,\%$ cholesterol, the number of water
molecules per lipid has increased by $4$. For this and higher cholesterol
concentrations the pressure in the bulk water phase indeed becomes zero, as can
be seen from Fig.~\ref{figGlobDruck}.

Second, the lateral pressure profile is not flat but some parts of the bilayer
would like to expand, at the same time that other parts would like to contract.
While the net pressure, averaged over the entire bilayer, is small, the local
pressure can be much higher than typical macroscopic pressures. A simple
estimate shows that the local lateral pressure in the bilayer core can be well
over $300~\mathrm{bar}$ and close to the interface can reach values exceeding
$1000~\mathrm{bar}$~\cite{gullingsrud:04a}. Our results for pure DPPC agree both
qualitatively and quantitatively with earlier results~\cite{lindahl:00b} with
small differences in the peak positions close the interface due to the different
treatment of electrostatics in that study.

Third, the magnitude of the local lateral pressure becomes higher as the
cholesterol concentration increases. While for small cholesterol concentrations
the local pressure is of the order of few hundred bars, it increases to
thousands of bar for the highest cholesterol concentrations. There is no
straightforward explanation for this but this phenomenon is very likely related
to higher bilayer rigidity at high cholesterol concentration since pressure
gradients are ultimately related to the elastic modulus.

Finally, the pressure profiles possess additional structure in the presence of
cholesterol. Without cholesterol, the lateral pressure in the lipid tail region
of the bilayer is strictly nonnegative. Already for $5\,\%$ cholesterol, a small
region of negative lateral pressure is seen in the figure. For cholesterol
concentrations of $20\,\%$ and higher, additional structure is seen. This
reflects that cholesterol is not some generic structureless object but rather
possesses an internal structure. One could call this the \emph{specific} effect
of cholesterol, compared to the \emph{unspecific} effects that are also
observed.

\begin{figure*}
\includegraphics[width=\columnwidth]{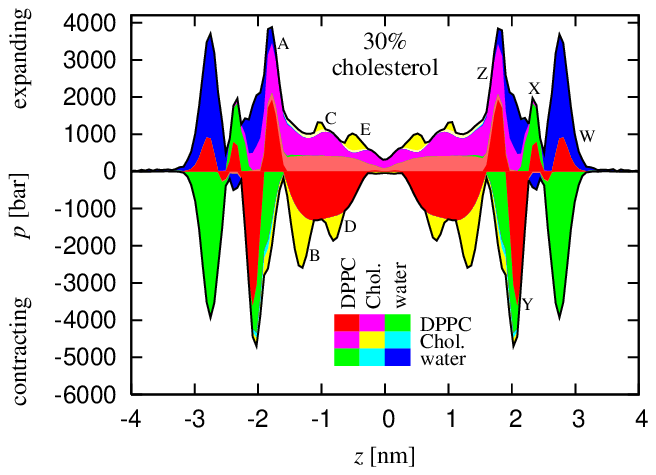}
\begin{picture}(210,100)
\put(10,10){\includegraphics[width=6.9cm]{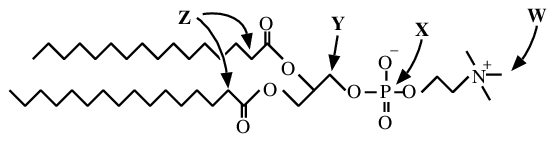}}
\put(10,77){\includegraphics[width=4.6cm]{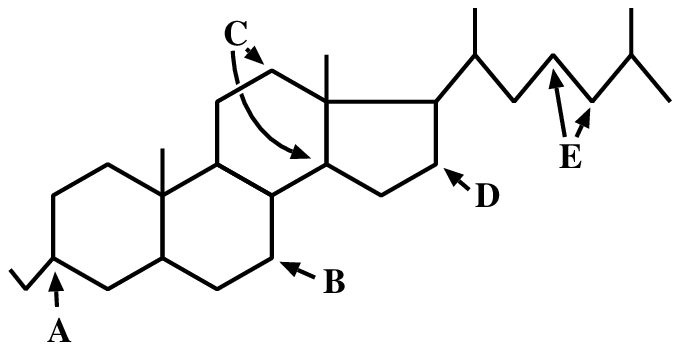}}
\put(180,70){DPPC}
\put(160,120){cholesterol}
\end{picture}\\
\includegraphics[width=\textwidth]{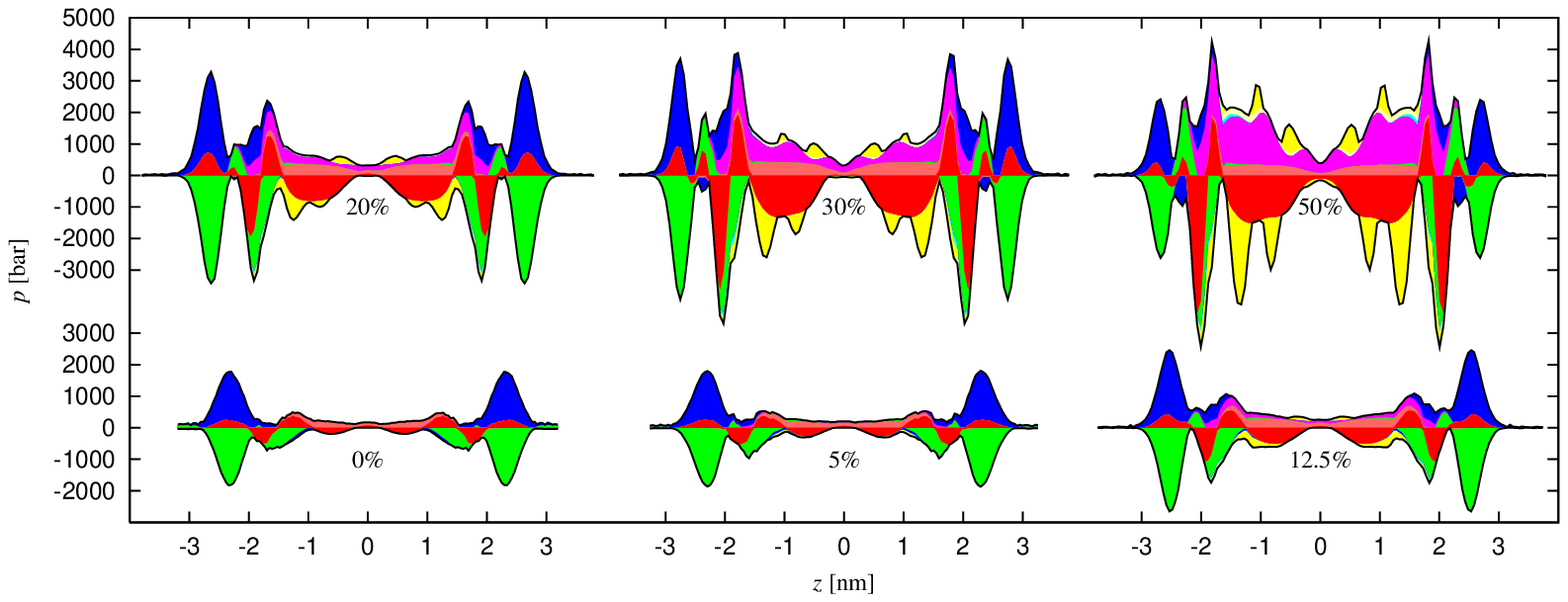}
\caption{Partial pressure profile through the bilayers. 
Expanding ($p>0$) and contracting ($p<0$) contributions are shown separately.
The colours code the origin of the pressure; lighter colours stand for
kinetic contributions. The label on the curves gives the cholesterol
concentration (bottom). On the top, an enlarged version of the profile for
$30\,\%$ cholesterol is shown where 
important peaks are
labelled according to which position in a DPPC and\,/\,or cholesterol molecule
they relate to. Schematic drawings of these molecules are included on the right.
The peaks for other cholesterol concentrations could not be labelled for
space reasons but they are easily mapped to the top figure.
}
\label{figPartial}
\end{figure*}

\section{Partial pressures}
\label{secPartial}

The pressure tensor~(\ref{eqDruck}) arises from inter- and intramolecular
pairwise forces [cf. $\tensor{\Sigma}$ from Eq.~(\ref{eqKraft})] as well as from
the kinetic motion of the atoms [cf. $\tensor{E}$ from Eq.~(\ref{eqEnergie})].
There are three components in the bilayer system (DPPC, cholesterol and water),
which gives six possible combinations for the pairwise forces, and
$\tensor{\Sigma}$ can thus be split into six contributions (DPPC--DPPC,
DPPC--cholesterol, and so on), according to which kind of molecules the two
atoms $i$ and $j$ causing the force $\vec{F}_{ij}$ belong to. 

Similarly, the kinetic energy tensor $\tensor{E}$ can be split into three
contributions. Thus, we divide the lateral pressure profile $p(z)$ into nine
contributions -- six from $\tensor{\Sigma}$ and three from $\tensor{E}$.  The
partial pressure profiles in Fig.~\ref{figPartial} show the different
contributions marked by different colours. The size of the coloured area
directly gives the pressure due to that contribution. The arrangement of the
colours, i.\,e., whether a given colour is close to the $z$-axis or further
away, has no physical meaning. The kinetic contribution from the DPPC molecules
is marked with the same colour tone but somewhat lighter (i.\,e., light red)
than the DPPC--DPPC contribution from the forces (i.\,e., dark red), and so on. 

There exists a fundamental difference between the tensor $\tensor{\Sigma}$, in
that it depends only on positions and forces and thus describes only the static
properties of the system, and $\tensor{E}$, which depends only on the velocities
and thus is a purely dynamic quantity. The kinetic energy per particle in 
Eq.~(\ref{eqEnergie}) can be larger than $k_{\mathrm{B}} T$ as, e.\,g., bond
vibrations or nonbonded interactions contribute energy to the translational
degrees of freedom. The
increase in kinetic pressure $\tensor{E}$ depends on whether these interactions
are mainly oriented parallel to the $x$-$y$ plane or normal to it. It thus
complements the traditional deuterium or NMR order  parameter
$|S_{\mathrm{CD}}|$ that is used to quantify orientational orientation along a
chain~\cite{tieleman:97a}.

Each contribution can be either expanding (positive pressure) or contracting
(negative pressure). The sum of all expanding contributions is shown as black
line in the figure. The same applies to the sum of all contracting 
contributions. The difference between these two values gives the pressure
profile that was shown in Fig.~\ref{figGlobDruck}. It should be noted that this
difference is significantly smaller in magnitude than each of the two original
terms.

As more cholesterol is added, the positions of
the pressure profile peaks shift. Simultaneous analysis of both the pressure
profile and the atom density profiles as a function of cholesterol concentration
allows us to find correlations between pressure and atom positions. The result
of this analysis is included in Fig.~\ref{figPartial}. Labels ``A''--``E''
refer to peak positions caused by cholesterol, and labels ``W''--``Z'' to DPPC.

Let us now take a closer look at the different regions in the bilayer.

\subsection{Interface region}

There is hardly any contact between water and cholesterol molecules, so one
might be tempted to assume that the interface region (also referred to
as headgroup region) of the bilayer should be hardly influenced
by cholesterol at all. The opposite is the case, as is seen from
Fig.~\ref{figPartial}. Cholesterol reduces the area per lipid of the bilayer,
and that strongly influences also the interface region, but the effects there
are mainly generic and unspecific -- any substance that would reduce the area
per lipid would have a similar effect.

The outermost part of the bilayer is formed by the choline group of the DPPC
molecules (labelled as ``W''), and the total pressure there is negative. The
attraction is dominated by the solvation energy between water and the  polar
lipid headgroups. Most of the the contracting pressure is compensated by the
positive pressure among the water molecules. Entropy plays an important role in
the latter as water molecules become ordered in the electrostatic field from the
zwitterionic headgroups~\cite{lindahl:00b}. The behaviour of this region of the
pressure profile depends only weakly on cholesterol concentration as the
hydration energy depends only weakly on the area per lipid~\cite{marsh:96a}. 
Note that both the  DPPC bilayer (red) and the water phase (blue) would like to
expand, and it is only the interaction between these two (green), commonly
called simply ``hydrophobicity'', that is trying to reduce the area of the
interface (green).

A bit further down in the bilayer, the phosphate groups of the DPPC molecules
are located. The mutual arrangement between the choline and phosphate groups
depends heavily on the area per lipid~\cite{gurtovenko:04a,falck:04a}. For large
area per lipid, and thus for low cholesterol concentration, both groups are, on
average, located in the same plane. In this case, the effects of the phosphate
and the choline group cannot be separated in the pressure profile. At
sufficiently large cholesterol concentration, the choline group is tilted up,
and an additional peak (``X'') thus appears. The pressure is positive as the
water molecules are oriented according to the charge of the choline group and
thus have an unfavourable interaction with the oppositely charged phosphate
group.

The inner boundary of the interface region is marked by the $sn-3$ carbon
(labelled as ``Y''). This is the
furthest extension to where there is noteworthy penetration of water into the
bilayer. Furthermore, this is the outermost position of cholesterol, occupied by
its hydroxyl group (``A''). Only inside this peak, cholesterol has, in addition
to its change of the area per lipid, also direct effects.

\subsection{Acyl chain region}

The acyl chain region is of special importance to the study of
cholesterol-containing bilayer as here not only the acyl chains of DPPC are
located (hence the name) but also the rigid four-ring structure of the
cholesterol body. In the presence of cholesterol, a characteristic structure in
the pressure profile develops, to the point that one might speculate on whether
this is sufficient to explain specific effects, such as observed for
the nicotinic acetylcholine receptor~\cite{rankin:97a}.

Upon increase of cholesterol concentration, an ordering of the DPPC acyl chains
takes place.  This ordering increases the attraction among the DPPC tails,
thereby increasing the compressing pressure component, while at the same time
introducing orientational correlations into the motion of the atoms of the
tails, thereby increasing the expanding kinetic pressure. The kinetic pressure
per DPPC molecule increases monotonously by almost a factor of $4$ from $0\,\%$
to $50\,\%$ cholesterol, and per cholesterol molecule it increases by a factor
of $3$. Cholesterol thus induces significant correlations into the motion of the
atoms, and the ordering induced by cholesterol is reflected much stronger in the
correlated motion of the atoms than it is in the  order parameter
$|S_{\mathrm{CD}}|$.

The rigid four-ring structure of cholesterol offers a much more interesting
chemical structure than the basically linear chains of the DPPC molecules. This
shows also in the pressure profile that exhibits several well-pronounced peaks,
making the pressure profiles of cholesterol-containing membranes very different
from previously reported pressure profiles~\cite{lindahl:00b,gullingsrud:04a}.
There are several atoms in the cholesterol molecules that are correlated with
peaks in the pressure profile, see the labels in Fig.~\ref{figPartial}. Steric
arguments can explain only same of the correlations. This should come as no
surprise since the pressure $p(z)$ at given depth $z$ in the bilayer cannot be
computed solely from atomic information from that depth -- the stresses in
different parts of the membrane are coupled due to finite elastic modulus. The
picture becomes even more difficult if one accounts also for chemistry, i.\,e.,
for favourable and unfavourable interactions between certain pairs of atoms.

The orientational correlations in the kinetic pressure, related to entropy and
giving rise to kinetic pressure, complicate matters even more. There is thus
little point in trying to explain the origin of every peak in the lateral
partial pressure profile, especially since at the moment there is little
possibility to prove or disprove any suggestion. What is sure, however, is that
all the features of the lateral pressure profile are in some way related to the
chemistry and structure of cholesterol.

Cholesterol is by far the most prominent steroid in any Eukaryote but it is only
one member of a large group of chemically related substances that are varying
only by seemingly minor changes in their structure, such as a single
additional double bound or a different placement of the two carbon atoms that
are sticking out from the rigid cholesterol body. Still, their properties in the
membrane are distinguishably different~\cite{endress:02a,scheidt:03a}, and only
in special cases one type of steroid can replace another in biological systems. 
Our results suggest that cholesterol relatives such as, e.\,g., lanosterol or
ergosterol, would have a pressure profile that is different from the pressure
profile of cholesterol systems. Unfortunately, no such simulations, let alone
experimental measurements, are available at the moment.

\subsection{Centre of the bilayer}

\begin{figure}
\centering
\includegraphics[height=7.2cm,trim=0 0 9 0,clip]{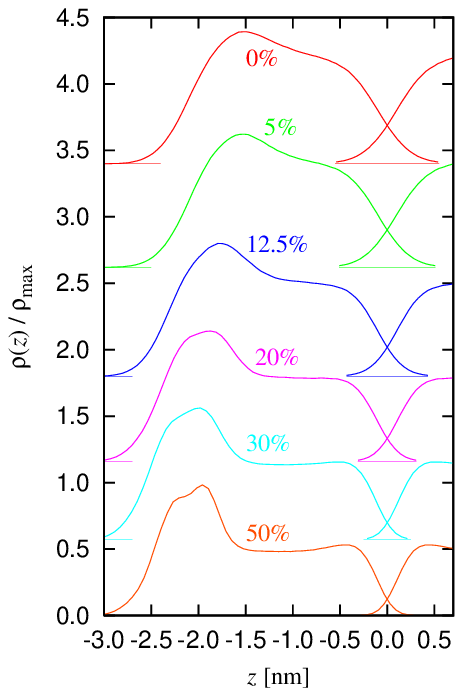}%
\includegraphics[height=7.2cm,trim=31 0 0 0,clip]{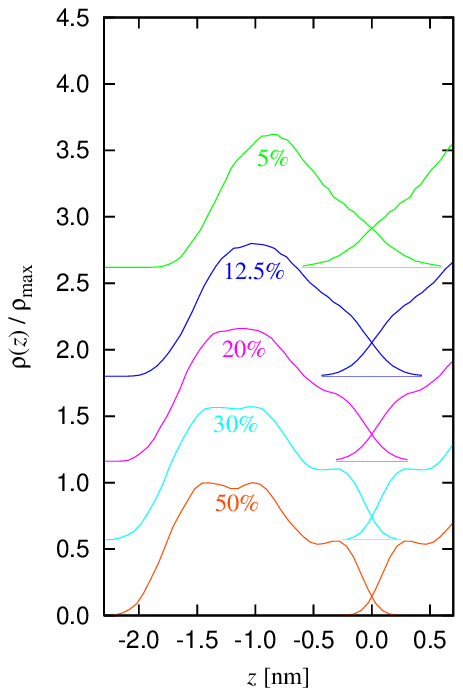}
\caption{Mass density profiles, with the contributions from both leaflets
plotted separately (left for DPPC, right for cholesterol). The curves have been
scaled by their maximum and shifted vertically.
For symmetry reasons,
only a small part of the right leaflet is shown. The numbers on the curves give
the cholesterol concentration. The amount of interdigitation is significantly
reduced if cholesterol is present.
}
\label{figInter}
\end{figure}

The centre of the bilayer is formed by the ends of the acyl chains of the DPPC
and cholesterol molecules. Both mass and electron density have a minimum at
$z=0$ known as methyl trough. For low cholesterol concentration, there is a
single peak in the pressure profile at $z=0$ that has been suggested to be
related to interdigitation~\cite{brink:04a}. Understanding the centre of the
bilayer is difficult as the peak in the pressure profile is found only in
atomistic MD simulations~\cite{lindahl:00b,gullingsrud:04a} whereas analytical
theories and coarse-grained models fail to reproduce it~\cite{brink:04a}. 
Experimentally it was found that cholesterol reduces
interdigitation but it should be noted that even pure DPPC bilayers 
are only moderately interdigitated~\cite{siminovitch:87a}.
This is in agreement with our results in Fig.~\ref{figInter}.

For increasing cholesterol concentration, the single peak in the pressure
profile at $z=0$ splits  into two peaks as the pressure close to $z=0$ increases
faster than precisely at $z=0$. Simultaneously, in addition 
to a reduction in interdigitation, the end point of the acyl chains
becomes more strongly defined pronounced in the mass density profile.
At the highest cholesterol concentrations, even a small peak in the mass 
density profiles in Fig.~\ref{figInter} develops near the end of the chains. 

The relation between interdigitation and pressure profile in the centre thus is
not obvious -- our results are more consistent with the absence of such a
relation. From
Fig.~\ref{figPartial} it is seen that the peak in the pressure profile 
around the bilayer centre does not
originate in an increase of the expanding pressure components but rather from a
sharp decrease of the condensing pressure. A possible explanation is that,
as they are less ordered than the other atoms, 
the atoms at the end of acyl chains cannot make favourable contact with neighbouring
chains, thereby loosing van der Waals-attraction. The shape of the pressure
profile near the centre of the bilayer would thus be dominated not by the mass
density profile itself but by the spatial distribution of end points of the 
chains. For cholesterol, this gives a peak at $z=0$ for small cholesterol
concentration that splits into two peaks as interdigitation is reduced, in
agreement with the computed pressure profiles.

\section{Relation to membrane penetration}

The lateral pressure profile is related to the ability of small molecules to
penetrate into the bilayer. Negative pressure means that the system would like
to contract. Regions with negative lateral pressure thus mark regions where it
would be energetically favourable for additional particles or molecules to be
inserted. Ultimately, the partitioning of a particular solute into the bilayer
is determined by the free-energy profile~\cite{marrink:96b}. The
lateral pressure profile gives the volume contribution to the free-energy
profile, and for nonpolar or moderately polar solutes, this is a good
approximation to the full free-energy profile.

On first view, this seems to be related to free-volume theories which state
that the amount of unoccupied space inside the bilayer determines its
penetrability. In the context of pressure profiles, ``occupied volume'' is
equivalent to large steric repulsive forces, resulting in a large expanding
static pressure. Already from the discussion in Sec.~\ref{secPartial} it should
be obvious that free-volume theories face a big problem as much of the pressure
inside the bilayer is either attractive or of kinetic origin.

One of the many reasons why cholesterol containing membranes are interesting
systems to study is that cholesterol affects the membrane spatially
inhomogeneously~\cite{lee:04a}. For example, the addition of cholesterol
increases the penetration of water into most regions of the bilayer, while it
reduces the water concentration in the very centre of the
bilayer~\cite{marsh:02a}. 


These observations are easily explained within the context of lateral pressure
profiles but not within free-volume theories. In the pure DPPC
bilayer, the pressure in the acyl chain region is small but positive. As the
cholesterol concentration is increased from $0\,\%$ to $30\,\%$, the pressure in
most of the acyl chain region becomes negative (allowing for easier penetration)
while in the centre of the bilayer it becomes more positive (thus expelling
particles there). Also at $50\,\%$ cholesterol, there are ample regions with
negative pressure even though they become disconnected by high peaks of positive
pressure. 

\begin{figure}
\includegraphics[width=0.8\columnwidth]{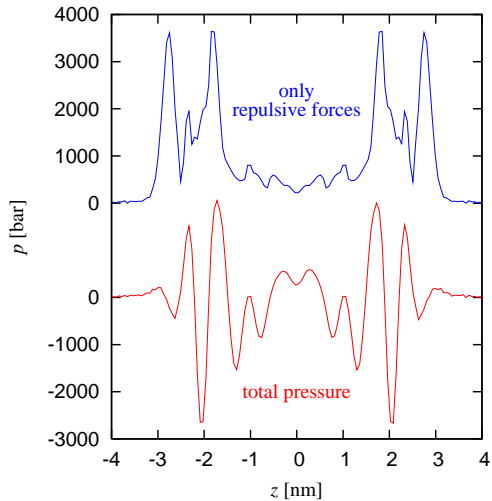}
\caption{Pressure profiles for $30\,\%$ cholesterol. The lower curve gives the
correct pressure profile whereas in the upper curve only static expanding
contributions to the pressure have been included.}
\label{figStatic}
\end{figure}

To make contact with free-volume theories,  it is instructive to ``compute''
from Fig.~\ref{figPartial} a ``fake'' pressure profile that only includes
static expanding pressure contributions, shown in Fig.~\ref{figStatic}.
The ``correct'' lateral-pressure profile changes strongly as
cholesterol is added, with several new peaks appearing in the pressure profile,
and the magnitude of the pressure changing by about one order of magnitude. In
contrast, the free-volume profile lacks the additional structure induced by
cholesterol and only changes monotonously as the cholesterol
concentration is changed, and the changes are rather moderate (at least compared
to the lateral-pressure profile)~\cite{falck:04a,falck:04b} -- both observations
apply also to the ``fake'' pressure profile.

The free-volume fraction is highest in the centre of the bilayer, which is
simply a reflection of the bilayer trough with reduced atom
density~\cite{falck:04b}, and molecules such as water should thus preferably be
located there. From this contradiction, it was thus realised already early that
it is not the average amount of free volume that determines the
penetration but rather rare fluctuations~\cite{marrink:94a}. 

Free-volume theory can be applied to compare penetration of solutes of different
size through the same bilayer, as penetration then is rate-limited by a single
point in the membrane, usually located at the bilayer
interface~\cite{killian:01a}. Already when comparing the same bilayer at
different temperatures, such a one-parameter approach may break
down~\cite{sutter:04a}. The lateral-pressure profile, on the other hand, allows
to compare also very different systems, such as bilayers with different
cholesterol concentration, as the description of the bilayer  is in full detail.
(The description of the solute is still reduced to solely its volume, and the
most straight-forward improvement to this simplification are free-energy
profiles.)

\section{Discussion}

In this paper we have studied lateral-pressure profiles in cholesterol--DPPC
bilayers at varying composition ratios. We found that the pressure profiles
become more structured and complicated as the cholesterol fraction is increased
(cf.~Fig.~\ref{figGlobDruck}).  This is related to the structural changes in the
bilayer upon addition of cholesterol. These changes can have  biological
consequences: proteins or other molecules are able to anchor at a given depth
inside a cell membrane guided by the pressure profile~\cite{brink:04a}.

Biological interactions can be divided into specific and unspecific. The general
view in the literature seems to be that changes via the lateral-pressure profile
are rather unspecific, i.\,e., proteins or other molecules in a cell membrane
are affected by the lateral-pressure profile but it is of secondary importance
of how a change in pressure-profile was induced. For specific interaction, a direct
contact between cholesterol and, e.\,g., a membrane channel, would be needed. 

In view of the lateral-pressure profiles presented in this paper one could
speculate whether a specific interaction via the pressure-profile is possible.
The profile contains a wealth of additional structure, very different from what
could be achieved by simply compressing the bilayer. This structure in the
lateral-pressure profile might be specific enough to allow molecules such as the
nicotinic acetylcholine receptor~\cite{rankin:97a} to sense the presence or
absence of cholesterol solely via the lateral-pressure profile 

This might be an additional reason for abundance of cholesterol in eukaryotic
cell membranes, and why there is so much cholesterol but so few other chemically
closely related steroids -- in contrast to the multitude of different lipids
found in cell membranes. From that point of view, it would be interesting to see
the difference between the different sterols on pressure profiles, e.\,g., how
the pressure profile would change if cholesterol was replaced by lanosterol.


The results presented in this paper also demonstrate that the effects of
cholesterol cannot be captured by simple models. Still, generic models have
their advantages, and the importance of lateral pressure profiles would not have
been accepted without the pioneering analytical work by Cantor in the late
1990's~\cite{cantor:97a,cantor:97b,cantor:99a,cantor:99b}. His most important
contribution properly were not so much the results themselves -- much of it was
already known -- but to cast them into a single consistent form that,
e.\,g., allowed to relate pressure profiles to ``nonbilayer lipids'' or
spontaneous membrane curvature.

Characteristic of such analytical theories is that they are very successful in
describing plain acyl chains but already their treatment of the headgroups is
rather crude~\cite{ben:95a,brink:04a}. Cholesterol is well beyond the
reach of such models. In the headgroup region, the problem is the description of
the mutual headgroup interactions and, even though this effect is generic,
their change upon decrease in the area per
lipid induced by cholesterol. In the acyl chain region, the problem is the
non-generic nature of the rigid cholesterol body.

This and other atomistic studies show the power of computer simulations to
resolve the details of molecular systems and to provide further insight to
complex multicomponent systems. Pressure profiles are a superb example for this
power due to the severe limitations encountered in experiment, coarse-grained
simulations and analytical theories.

\acknowledgments

I would like to acknowledge valuable discussions with Mikko Karttunen 
and support from the European Union (MRTN-CT-2004-512331).


\begin{thebibliography}{56}
\providecommand{\natexlab}[1]{#1}

\bibitem[{An{\'e}zo et~al.(2003)An{\'e}zo, de~Vries, H{\"o}ltje, Tieleman, and
  Marrink}]{anezo:03a}
An{\'e}zo, C., A.~H. de~Vries, H.-D. H{\"o}ltje, D.~P. Tieleman, and S.-J.
  Marrink. 2003.
\newblock Methodological issues in lipid bilayer simulations.
\newblock \emph{J. Phys. Chem. B} 107:9424--9433.

\bibitem[{Ben-Shaul(1995)}]{ben:95a}
Ben-Shaul, A. 1995.
\newblock Molecular theory of chain packing, elasticity and lipid-protein
  interaction in lipid bilayers.
\newblock vol.~1 of \emph{Handbook of Biological Physics}. Elsevier, Amsterdam,
  359--401.

\bibitem[{Berendsen et~al.(1984)Berendsen, Postma, van Gunsteren, {Di Nola},
  and Haak}]{berendsen:84a}
Berendsen, H. J.~C., J.~P.~M. Postma, W.~F. van Gunsteren, A.~{Di Nola}, and
  J.~R. Haak. 1984.
\newblock Molecular dynamics with coupling to an external bath.
\newblock \emph{J. Chem. Phys.} 81:3684--3690.

\bibitem[{Berendsen et~al.(1981)Berendsen, Postma, van Gunsteren, and
  Hermans}]{berendsen:81a}
Berendsen, H. J.~C., J.~P.~M. Postma, W.~F. van Gunsteren, and J.~Hermans.
  1981.
\newblock Interaction models for water in relation to protein hydration.
\newblock In Intermolecular Forces, B.~Pullman, editor. Reidel, Dordrecht,
  331--342.

\bibitem[{Berger et~al.(1997)Berger, Edholm, and Jahnig}]{berger:97a}
Berger, O., O.~Edholm, and F.~Jahnig. 1997.
\newblock Molecular dynamics simulations of a fluid bilayer of
  dipalmitoylphosphatidylcholine at full hydration, constant pressure, and
  constant temperature.
\newblock \emph{Biophys. J.} 72:2002--2013.
\newblock The force field description is available at
  http://moose.bio.ucalgary.ca/Downloads/files/lipid.itp.

\bibitem[{Bezrukov(2000)}]{bezrukov:00a}
Bezrukov, S.~M. 2000.
\newblock Functional consequences of lipid packing stress.
\newblock \emph{Curr. Opin. Colloid Interface Sci.} 5:237--243.

\bibitem[{Bishop et~al.(1997)Bishop, Skeel, and Schulten}]{bishop:97a}
Bishop, T.~C., R.~D. Skeel, and K.~Schulten. 1997.
\newblock Difficulties with multiple time stepping and fast multipole algorithm
  in molecular dynamics.
\newblock \emph{J. Comput. Chem.} 18:1785--1791.

\bibitem[{Cantor(1997{\natexlab{a}})}]{cantor:97b}
Cantor, R.~S. 1997{\natexlab{a}}.
\newblock The lateral pressure profile in membranes: a physical mechanism of
  general anesthesia.
\newblock \emph{Biochemistry} 36:2339--2344.

\bibitem[{Cantor(1997{\natexlab{b}})}]{cantor:97a}
Cantor, R.~S. 1997{\natexlab{b}}.
\newblock Lateral pressures in cell membranes: a mechanism for modulation of
  protein function.
\newblock \emph{J. Phys. Chem.} 101:1723--1725.

\bibitem[{Cantor(1999{\natexlab{a}})}]{cantor:99b}
Cantor, R.~S. 1999{\natexlab{a}}.
\newblock The influence of membrane lateral pressures on simple geometric
  models of protein conformational equilibria.
\newblock \emph{Chem. Phys. Lipids} 101:45--56.

\bibitem[{Cantor(1999{\natexlab{b}})}]{cantor:99a}
Cantor, R.~S. 1999{\natexlab{b}}.
\newblock Lipid composition and the lateral pressure profile in bilayers.
\newblock \emph{Biophys. J.} 76:2625--2639.

\bibitem[{de~Kruijff(1997)}]{kruijff:97a}
de~Kruijff, B. 1997.
\newblock Lipids beyond the bilayer.
\newblock \emph{Nature} 386:129--130.

\bibitem[{Eckenhoff(2001)}]{eckenhoff:01a}
Eckenhoff, R.~G. 2001.
\newblock Promiscuous ligands and attractive cavities: How do the inhaled
  anesthetics work?
\newblock \emph{Mol Interv.} 1:258--268.

\bibitem[{Endress et~al.(2002)Endress, Heller, Casalta, Brown, and
  Bayerl}]{endress:02a}
Endress, E., H.~Heller, H.~Casalta, M.~F. Brown, and T.~M. Bayerl. 2002.
\newblock Anisotropic motion and molecular dynamics of cholesterol, lanosterol,
  and ergosterol in lecithin bilayers studied by quasi-elastic neutron
  scattering.
\newblock \emph{Biochemistry} 41:13078--13086.

\bibitem[{Essmann et~al.(1995)Essmann, Perera, Berkowitz, Darden, Lee, and
  Pedersen}]{essman:95a}
Essmann, U., L.~Perera, M.~L. Berkowitz, T.~Darden, H.~Lee, and L.~G. Pedersen.
  1995.
\newblock A smooth particle mesh {E}wald method.
\newblock \emph{J. Chem. Phys.} 103:8577--8592.

\bibitem[{Falck et~al.(2004{\natexlab{a}})Falck, Patra, Karttunen, Hyv{\"o}nen,
  and Vattulainen}]{falck:04b}
Falck, E., M.~Patra, M.~Karttunen, M.~T. Hyv{\"o}nen, and I.~Vattulainen.
  2004{\natexlab{a}}.
\newblock Impact of cholesterol on voids in phospholipid membranes.
\newblock \emph{J. Chem. Phys.} 121:12676--12689.

\bibitem[{Falck et~al.(2004{\natexlab{b}})Falck, Patra, Karttunen, Hyv{\"o}nen,
  and Vattulainen}]{falck:04a}
Falck, E., M.~Patra, M.~Karttunen, M.~T. Hyv{\"o}nen, and I.~Vattulainen.
  2004{\natexlab{b}}.
\newblock Lessons of slicing membranes: Interplay of packing, free area, and
  lateral diffusion in phospholipid/cholesterol bilayers.
\newblock \emph{Biophys. J.} 87:1076--1091.

\bibitem[{Frenkel and Smit(2002)}]{frenkel:02}
Frenkel, D. and B.~Smit. 2002.
\newblock Understanding Molecular Simulation: From Algorithms to Applications.
\newblock Academic Press, San Diego, second ed.

\bibitem[{Goetz and Lipowsky(1998)}]{goetz:98a}
Goetz, R. and R.~Lipowsky. 1998.
\newblock Computer simulations of bilayer membranes: Self-assembly and
  interfacial tension.
\newblock \emph{J. Chem. Phys.} 108:7397--7409.

\bibitem[{Gullingsrud and Schulten(2004)}]{gullingsrud:04a}
Gullingsrud, J. and K.~Schulten. 2004.
\newblock Lipid bilayer pressure profiles and mechanosensitive channel gating.
\newblock \emph{Biophys. J.} 86:3496--3509.

\bibitem[{Gurtovenko et~al.(2004)Gurtovenko, Patra, Karttunen, and
  Vattulainen}]{gurtovenko:04a}
Gurtovenko, A.~A., M.~Patra, M.~Karttunen, and I.~Vattulainen. 2004.
\newblock Cationic {DMPC}/{DMTAP} lipid bilayers: Molecular dynamics study.
\newblock \emph{Biophys. J.} 86:3461--3472.

\bibitem[{Hamill and Martinac(2001)}]{hamill:01a}
Hamill, O.~P. and B.~Martinac. 2001.
\newblock Molecular basis of mechanotransduction in living cells.
\newblock \emph{Physiol. Rev.} 81:685--740.

\bibitem[{Harries and Ben-Shaul(1997)}]{harries:97a}
Harries, D. and A.~Ben-Shaul. 1997.
\newblock Conformational chain statistics in a model lipid bilayer: Comparison
  between mean field and {M}onte {C}arlo calculations.
\newblock \emph{J. Chem. Phys.} 106:1609--1619.

\bibitem[{Heinz et~al.(2004)Heinz, Paul, and Binder}]{heinz:03a}
Heinz, H., W.~Paul, and K.~Binder. 2004.
\newblock Local pressure tensor in computer simulations of molecular systems.
  arXiv.org:cond-mat/0309014.

\bibitem[{Hess et~al.(1997)Hess, Bekker, Berendsen, and Fraaije}]{hess:97a}
Hess, B., H.~Bekker, H.~J.~C. Berendsen, and J.~G. E.~M. Fraaije. 1997.
\newblock {LINCS}: A linear constraint solver for molecular simulations.
\newblock \emph{J. Comp. Chem.} 18:1463--1472.

\bibitem[{Hofs{\"a}{\ss} et~al.(2003)Hofs{\"a}{\ss}, Lindahl, and
  Edholm}]{hofsaess:03a}
Hofs{\"a}{\ss}, C., E.~Lindahl, and O.~Edholm. 2003.
\newblock Molecular dynamics simulations of phospholipid bilayers with
  cholesterol.
\newblock \emph{Biophys. J.} 84:2192--2206.

\bibitem[{H{\"o}ltje et~al.(2001)H{\"o}ltje, F{\"o}rster, Brandt, Engels, von
  Rybinski, and H{\"o}ltje}]{hoeltje:01a}
H{\"o}ltje, M., T.~F{\"o}rster, B.~Brandt, T.~Engels, W.~von Rybinski, and
  H.-D. H{\"o}ltje. 2001.
\newblock Molecular dynamics simulations of stratum corneum lipid models: fatty
  acids and cholesterol.
\newblock \emph{Biochim. Biophys. Acta} 1511:156--167.
\newblock The topology file is available from
  http://www.gromacs.org/topologies/uploaded\_molecules/
  cholesterol.tgz.

\bibitem[{Killian and van Meer(2001)}]{killian:01a}
Killian, J.~A. and G.~van Meer. 2001.
\newblock The `double lives' of membrane lipids.
\newblock \emph{EMBO reports} 21:91--95.

\bibitem[{Kinnunen(2000)}]{kinnunen:00a}
Kinnunen, P. K.~J. 2000.
\newblock Lipid bilayers as osmotic response elements.
\newblock \emph{Cell. Physiol. Biochem.} 10:243--250.

\bibitem[{Lee and Petersen(2004)}]{lee:04a}
Lee, C.~C. and N.~O. Petersen. 2004.
\newblock The triple layer model: A different perspective on lipid bilayers.
\newblock \emph{J. Chin. Chem. Soc.} 51:1183--1191.

\bibitem[{Lindahl and Edholm(2000)}]{lindahl:00b}
Lindahl, E. and O.~Edholm. 2000.
\newblock Spatial and energetic-entropic decomposition of surface tension in
  lipid bilayers from molecular dynamics simulations.
\newblock \emph{J. Chem. Phys.} 113:3882--3893.

\bibitem[{Lindahl et~al.(2001)Lindahl, Hess, and van~der Spoel}]{lindahl:01a}
Lindahl, E., B.~Hess, and D.~van~der Spoel. 2001.
\newblock {GROMACS} 3.0: a package for molecular simulation and trajectory
  analysis.
\newblock \emph{Journal of Molecular Modeling} 7:306--317.

\bibitem[{Marrink and Berendsen(1994)}]{marrink:94a}
Marrink, S.-J. and H.~J.~C. Berendsen. 1994.
\newblock Simulation of water transport through a lipid membrane.
\newblock \emph{J. Phys. Chem.} 98:4155--4168.

\bibitem[{Marrink and Berendsen(1996)}]{marrink:96b}
Marrink, S.~J. and H.~J.~C. Berendsen. 1996.
\newblock Permeation process of small molecules across lipid membranes studied
  by molecular dynamics simulations.
\newblock \emph{J. Phys. Chem.} 100:16729--16738.

\bibitem[{Marsh(1996)}]{marsh:96a}
Marsh, D. 1996.
\newblock Lateral pressure in membranes.
\newblock \emph{Biochim. Biophys. Acta} 1286:183--223.

\bibitem[{Marsh(2002)}]{marsh:02a}
Marsh, D. 2002.
\newblock Membrane water-penetration profiles from spin labels.
\newblock \emph{Eur. Biophys. J.} 31:559--562.

\bibitem[{McMullen and McElhaney(1996)}]{mcmullen:96a}
McMullen, D. P.~W. and R.~N. McElhaney. 1996.
\newblock Physical studies cholesterol-phospholipid interactions.
\newblock \emph{Curr. Opin. Colloid Interface Sci.} 1:83--90.

\bibitem[{Miyamoto and Kollman(1992)}]{miyamoto:92a}
Miyamoto, S. and P.~A. Kollman. 1992.
\newblock {SETTLE}: An analytical version of the {SHAKE} and {RATTLE}
  algorithms for rigid water models.
\newblock \emph{J. Comput. Chem.} 13:952--962.

\bibitem[{Patra et~al.(2003)Patra, Karttunen, Hyv{\"o}nen, Falck, Lindqvist,
  and Vattulainen}]{patra:03b}
Patra, M., M.~Karttunen, M.~Hyv{\"o}nen, E.~Falck, P.~Lindqvist, and
  I.~Vattulainen. 2003.
\newblock Molecular dynamics simulations of lipid bilayers: Major artifacts due
  to truncating electrostatic interactions.
\newblock \emph{Biophys. J.} 84:3636--3645.

\bibitem[{Patra et~al.(2004{\natexlab{a}})Patra, Karttunen, Hyv{\"o}nen, Falck,
  and Vattulainen}]{patra:04b}
Patra, M., M.~Karttunen, M.~T. Hyv{\"o}nen, E.~Falck, and I.~Vattulainen.
  2004{\natexlab{a}}.
\newblock Lipid bilayers driven to a wrong lane in molecular dynamics
  simulations by subtle changes in long-range electrostatic interactions.
\newblock \emph{J. Phys. Chem. B} 108:4485--4494.

\bibitem[{Patra et~al.(2004{\natexlab{b}})Patra, Karttunen, Hy{v\"o}nen, Falck,
  and Vattulainen}]{patra:04c}
Patra, M., M.~Karttunen, M.~T. Hy{v\"o}nen, E.~Falck, and I.~Vattulainen.
  2004{\natexlab{b}}.
\newblock Long-range interactions in molecular simulations: Accuracy and speed.
  arXiv.org:cond-mat/0410210.

\bibitem[{Rankin et~al.(1997)Rankin, Addona, Kloczewiak, Bugge, and
  Miller}]{rankin:97a}
Rankin, S.~E., G.~H. Addona, M.~A. Kloczewiak, B.~Bugge, and K.~W. Miller.
  1997.
\newblock The cholesterol dependance of activation and fast desensitization of
  the nicotinic acetylcholine receptor.
\newblock \emph{Biophys. J.} 73:2446--2455.

\bibitem[{Ryckaert et~al.(1977)Ryckaert, Ciccotti, and
  Berendsen}]{ryckaert:77a}
Ryckaert, J.-P., G.~Ciccotti, and H.~J.~C. Berendsen. 1977.
\newblock Numerical integration of the cartesian equations of motion of a
  system with constraints; molecular dynamics of n-alkanes.
\newblock \emph{J. Comp. Phys.} 23:327--341.

\bibitem[{Scheidt et~al.(2003)Scheidt, M{\"u}ller, Herrmann, and
  Huster}]{scheidt:03a}
Scheidt, H.~A., P.~M{\"u}ller, A.~Herrmann, and D.~Huster. 2003.
\newblock The potential of fluorescent and spin-labeled steroid analogs to
  mimic natural cholesterol.
\newblock \emph{J. Biol. Chem.} 278:45563--45569.

\bibitem[{Shillcock and Lipowsky(2002)}]{shillcock:02a}
Shillcock, J.~C. and R.~Lipowsky. 2002.
\newblock Equilibrium structure and lateral stress distribution of amphiphilic
  bilayers from dissipative particle dynamics simulations.
\newblock \emph{J. Chem. Phys.} 117:5048--5061.

\bibitem[{Siminovitch et~al.(1987)Siminovitch, Ruocco, Makriyannis, and
  Griffin}]{siminovitch:87a}
Siminovitch, D.~J., M.~J. Ruocco, A.~Makriyannis, and R.~G. Griffin. 1987.
\newblock The effect of cholesterol on lipid dynamics and packing in diether
  phosphatidylcholine bilayers. {X}-ray diffraction and {$^2$H-NMR} study.
\newblock \emph{Biochim. Biophys. Acta} 901:191--200.

\bibitem[{Simons and Ikonen(2000)}]{simons:00a}
Simons, K. and E.~Ikonen. 2000.
\newblock How cells handle cholesterol.
\newblock \emph{Science} 290:1721--1726.

\bibitem[{Sukharev et~al.(1997)Sukharev, Blount, Martinac, and
  Kung}]{suhkarev:97a}
Sukharev, S.~I., P.~Blount, B.~Martinac, and C.~Kung. 1997.
\newblock Mechanosensitive channels of {\em {e}scheria coli} -- the {MscL}
  gene, protein and activities.
\newblock \emph{Annu. Rev. Physiol.} 59:633--657.

\bibitem[{Sutter et~al.(2004)Sutter, Fiechter, and Imanidis}]{sutter:04a}
Sutter, M., T.~Fiechter, and G.~Imanidis. 2004.
\newblock Correlation of membrane order and dynamics derived from time-resolved
  fluorescence measurements with solute permeability.
\newblock \emph{J. Pharmaceutical Sciences} 93.

\bibitem[{Templer et~al.(1998)Templer, Castle, Curran, Rumbles, and
  Klug}]{templer:98a}
Templer, R.~H., S.~J. Castle, A.~R. Curran, G.~Rumbles, and D.~R. Klug. 1998.
\newblock Sensing isothermal changes in the lateral pressure in model membranes
  using di-pyrenyl phosphatidylchlonine.
\newblock \emph{Faraday Discuss.} 111:41--53.

\bibitem[{Tieleman and Berendsen(1996)}]{tieleman:96a}
Tieleman, D.~P. and H.~J.~C. Berendsen. 1996.
\newblock Molecular dynamics simulations of a fully hydrated
  dipalmitoylphosphatidylcholine bilayer with different macroscopic boundary
  conditions and parameters.
\newblock \emph{J. Chem. Phys.} 105:4871--4880.
\newblock The topology file is available from
  http://moose.bio.ucalgary.ca/Downloads/files/dppc.itp.

\bibitem[{Tieleman et~al.(1997)Tieleman, Marrink, and Berendsen}]{tieleman:97a}
Tieleman, D.~P., S.-J. Marrink, and H.~J.~C. Berendsen. 1997.
\newblock A computer perspective of membranes: Molecular dynamics studies of
  lipid bilayer systems.
\newblock \emph{Biochim. Biophys. Acta} 1331:235--270.

\bibitem[{Tironi et~al.(1995)Tironi, Sperb, Smith, and van
  Gunsteren}]{tironi:95a}
Tironi, I.~G., R.~Sperb, P.~E. Smith, and W.~F. van Gunsteren. 1995.
\newblock A generalized reaction field method for molecular dynamics
  simulations.
\newblock \emph{J. Chem. Phys.} 102:5451--5459.

\bibitem[{Tu et~al.(1998)Tu, Klein, and Tobias}]{tu:98a}
Tu, K., M.~L. Klein, and D.~J. Tobias. 1998.
\newblock Constant-pressure molecular dynamics investigation of cholesterol
  effects in a dipalmitoylphosphatidylcholine bilayer.
\newblock \emph{Biophys. J.} 75:2147--2156.

\bibitem[{van den Brink-van~der Laan et~al.(2004)van den Brink-van~der Laan,
  Killian, and de~Kruijff}]{brink:04a}
van den Brink-van~der Laan, E., J.~A. Killian, and B.~de~Kruijff. 2004.
\newblock Nonbilayer lipids affect peripheral and integral membrane proteins
  via changes in the lateral pressure profile.
\newblock \emph{Biochim. Biophys. Acta} 1666:275--288.

\bibitem[{Yeagle(1985)}]{yeagle:85a}
Yeagle, P.~L. 1985.
\newblock Cholesterol and the cell membrane.
\newblock \emph{Biochim. Biophys. Acta} 822:267--287.

\end{thebibliography}

\end{document}